# The quantum superposition principle: a reconsideration

## I. G. Koprinkov

Department of Applied Physics, Faculty of Applied Mathematics and Informatics, Technical University of Sofia, e-mail: igk@tu-sofia.bg

*Abstract: The quantum superposition principle is reconsidered based on adiabatic theorem of quantum mechanics, nonadiabatic dressed states and experimental evidence. The physical mechanism and physical properties of the quantum superposition are revealed.*

**Keywords:** quantum superposition, adiabatic theorem, dressed states.

## 1. Introduction

The quantum superposition principle is one of the fundamental principles of quantum mechanics. It states that if two or more quantum states $|\psi(\vec{r}, t)\rangle_i$ are states of a given quantum system, *i.e.*, they obey the Schrödinger equation, any linear combination of these states is also a state of that quantum system:

$$|\psi(\vec{r}, t)\rangle = \sum_i c_i(t) |\psi(\vec{r}, t)\rangle_i \qquad (1)$$

It is correct from a formal mathematical point of view due to the linearity of the Schrödinger equation. The following understandings are automatically assigned to superposition (1): the superposition of quantum states is *simultaneous*, *coherent* and may involve *any kind of states* of the quantum system. It means that the quantum system is simultaneously, but not consequently, in all superimposed states. Such simultaneity has never been proved experimentally. The typical characteristic time of electron motion in atoms falls in the attosecond time domain, $1as=10^{-18}s$. The present attophysics and technology is able to reach this time range but, to the best of our knowledge, the exact timing of quantum superposition has not been tested experimentally. The coherence means that the superimposed states maintain definite phase relation. Finally, the superimposed states can be any stationary eigenstates of quantum systems as atoms, ions, molecules, etc. In addition, the physical mechanism of this superposition is not specified assuming that is a natural consequence from the nature of the quantum phenomena. The aim of this work is to find *the physical mechanism* (instead of *mathematical formalism*) of quantum superposition, the physical features of the superimposed quantum states and the physical consequences from it as the collapse of wave function, the quantum measurement problem, etc. Our approach is based on adiabatic theorem of quantum mechanics [1], experimental studies on real and virtual quantum states [2, 3] and nonadiabatic dressed states [4].

## 2. Physical grounds in the formulation of quantum superposition principle

The quantum superposition principle will be formulated here based on the following physical grounds:
    1. General theoretical ground – adiabatic theorem of quantum mechanics [1].
    2. Special theoretical ground – nonadiabatic dressed states [4].
    3. Experimental evidences [2, 3].

### 2.1. Adiabatic theorem of quantum mechanics and quantum superposition

*The adiabatic theorem of quantum mechanics* [1] states that a quantum system remains in an instantaneous eigenstate $\psi(t)$ of its Hamiltonian $\hat{H}(t) = \hat{H}_0 + \hat{H}'(t)$, *i.e.*, $\hat{H}(t)\psi(t) \equiv (\hat{H}_0 + \hat{H}'(t))\psi(t) = E(t)\psi(t)$, if the latter changes slow enough, *i.e.*, adiabatically, due to given perturbation $\hat{H}'(t)$, and if there is a gap $\Delta E$ between the energy of this state and the rest part of the Hamiltonian spectrum, fig.1. When the perturbation terminates, the quantum system will be in the same quantum state $\psi_0$, from which the adiabatic evolution begins, and no transition to other state $\psi_0'$ will occur. Consequently, according to adiabatic theorem, *the quantum system cannot be simultaneously in more than one eigenstate of adiabatically changing Hamiltonian.*

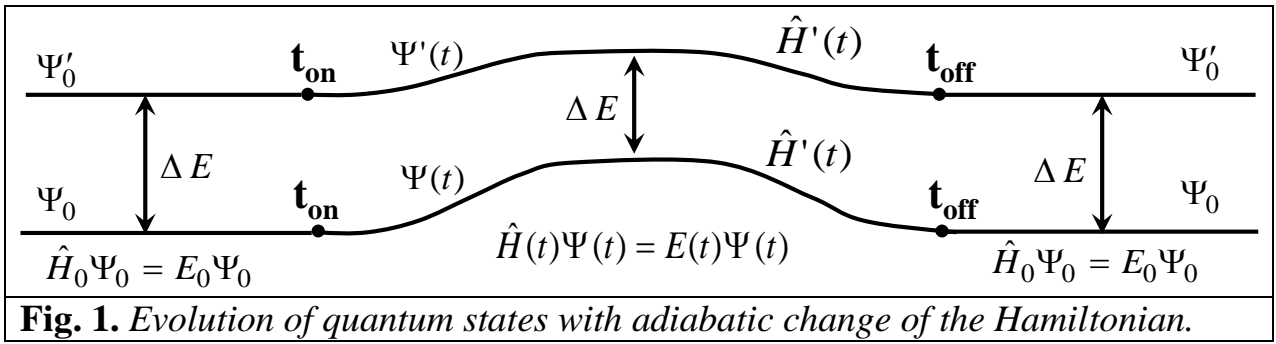

**Fig. 1.** *Evolution of quantum states with adiabatic change of the Hamiltonian.*

The adiabatic theorem can also be applied to the particular case of stationary eigenstates $\psi_0$ of the nonperturbed Hamiltonian $\hat{H}_0$ of a completely isolated, closed, quantum system, $\hat{H}_0 \psi_0 = E_0 \psi_0$. These states are known as bare states (BS). The BS represent states at perfect adiabatic conditions because the perturbation Hamiltonian $\hat{H}'(t)$ is not simply adiabatic but stationary, $\hat{H}'(t) = 0 = const$. Thus, the adiabatic theorem in this asymptotic adiabatic case states that *if a quantum system is in given BS, it will remain in that BS and no transition to other BS will occur*. Consequently, according to adiabatic theorem, *the quantum system cannot be simultaneously in more than one stationary eigenstate, i.e., BS, of the nonperturbed Hamiltonian $\hat{H}_0$*.

In agreement with the adiabatic theorem, the physical reasons for quantum transition are the nonadiabatic factors acting on the quantum system. The nonadiabatic factors can be put into two categories: "regular" nonadiabatic factors and stochastic nonadiabatic factors. To the first type belong rapid, nonadiabatic, "regular" variations of the electromagnetic field. To the second type belong stochastic variations of the electromagnetic field, collisions with other atoms, ions, molecules, zero-point vacuum fluctuations, etc. The nonadiabatic factors can be reduced but cannot be completely eliminated due to, at least, the zero-point vacuum fluctuations.

### 2.2. Nonadiabatic dressed states and quantum superposition

One may distinguish three generations of quantum states for a given quantum system: bare states (BS), adiabatic (dressed) states (ADS) [2, 3] and nonadiabatic dressed states (NADS) [4]. The BS are states of completely *isolated* (*closed*) *quantum system*. The ADS are states of quantum system in presence of classical adiabatic (slowly varying) electromagnetic field. These states are originally called *adiabatic states*, but as they are considered equivalent to their full quantum analog, they will be called here ADS. Finally, the NADS are states of *open quantum system* in presence of nonadiabatic factors from the electromagnetic field and the environment (collisions with other quantum systems, zero-point vacuum fluctuations, etc.). The following notations for the ground and the excited BS, ADS, NADS will be used here: $|g\rangle$ and $|e\rangle$, $|G\rangle$ and $|E\rangle$, $|\widetilde{G}\rangle$ and $|\widetilde{E}\rangle$, respectively. The NADS are a generalization of the ADS and the BS. The NADS and the ADS have same structure but as the NADS are only the states that include explicitly nonadiabatic factors, the NADS picture will be used here. The NADS and the relevant physical processes will be considered below.

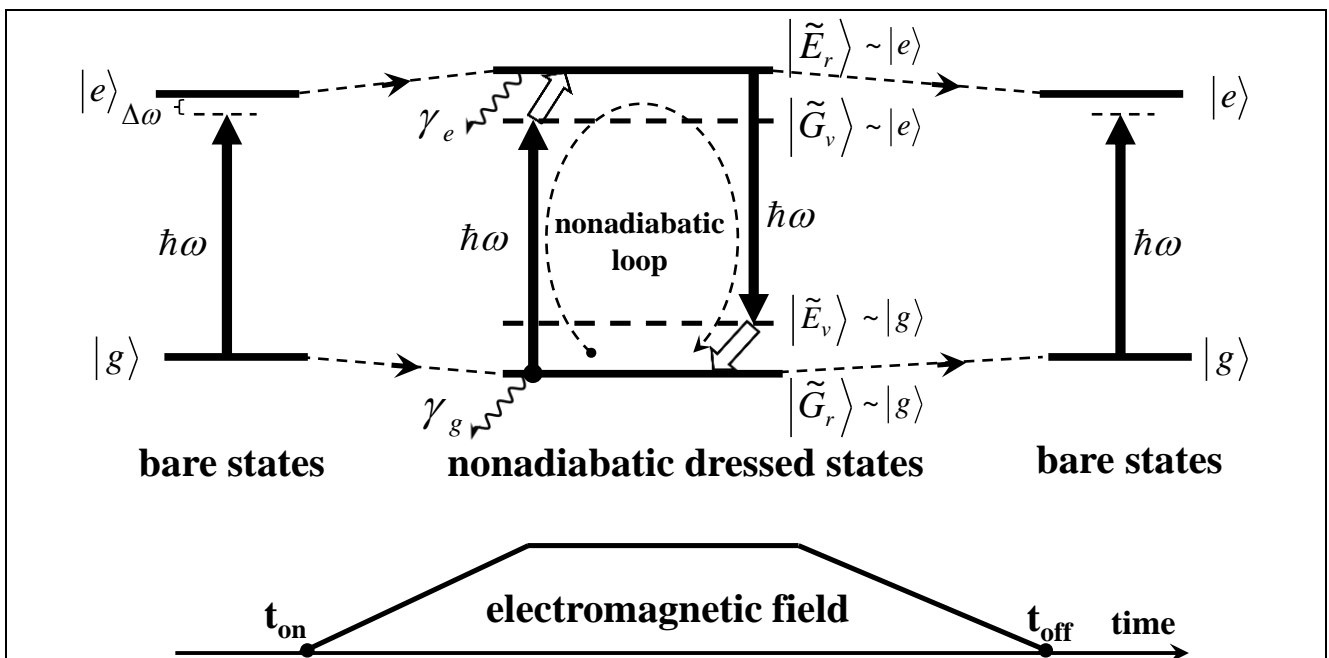

**Fig. 2.** *Evolution of BS toward NADS and back to BS with switching on/off the electromagnetic field and damping. The real and the virtual components of NADS are shown with full and broken lines, respectively. Solid arrows show radiative transitions, hollow arrows show nonadiabatic transitions.*

Consider an open two-level quantum system of electric dipole allowed transitions between initial BS, ground $|1\rangle \equiv |g\rangle$ and excited $|2\rangle \equiv |e\rangle$, Fig.2. The states will be called, for short, $|g\rangle$-type and $|e\rangle$-type, which includes all characteristics of the states (quantum numbers, symmetries, etc.). The NADS are constructed from an analytic solution of Schrödinger equation $\hat{H}|\Psi(\vec{r},t)\rangle = i\hbar\partial_t|\Psi(\vec{r},t)\rangle$ with Hamiltonian $\hat{H}$:

$$\hat{H} = \hat{H}_0 + \hat{H}' + \hat{H}_D = \sum_{j=1}^{2}\hbar\omega_j|j\rangle\langle j| - \mu E(t)\big(|1\rangle\langle 2| + h.c.\big) - i\hbar\sum_{j=1}^{2}\big(\gamma_j/2\big)|j\rangle\langle j| \quad (2)$$

The total Hamiltonian $\hat{H}$ consists of Hamiltonian of the quantum system itself $\hat{H}_0$, Hamiltonian of interaction of the quantum system with external electromagnetic field $\hat{H}'$ and Hamiltonian of interaction of quantum system with the environment $\hat{H}_D$, described by damping rates $\gamma_j$ [4]. The quantum system is subject to a "regular" interaction with an external nonadiabatic electromagnetic field and stochastic interactions with the environment (collisions, zero-point vacuum fluctuations, etc.).

Each NADS consists of real (index "$r$") and virtual (index "$v$") components [4]:

$$|\widetilde{G}\rangle = COS(\theta/2)|\widetilde{G}_r\rangle + SIN(\theta/2)|\widetilde{G}_v\rangle \quad (3a)$$

$$|\widetilde{E}\rangle = COS(\theta/2)|\widetilde{E}_r\rangle - SIN(\theta/2)|\widetilde{E}_v\rangle \quad (3b)$$

The real and the virtual components of NADS (at ground state initial conditions) are:

$$|\widetilde{G}_r\rangle = |g\rangle \exp\left[-i\int_0^t \widetilde{\omega}_G dt'\right] \quad (4a)$$

$$|\widetilde{G}_v\rangle = |e\rangle \exp\left[-i\int_0^t (\widetilde{\omega}_G + \omega)dt' - i\varphi(t)\right] \quad (4b)$$

$$|\widetilde{E}_r\rangle = |e\rangle \exp\left[-i\int_0^t \widetilde{\omega}_E dt' - i\varphi(t)\right] \quad (4c)$$

$$|\widetilde{E}_v\rangle = |g\rangle \exp\left[-i\int_0^t (\widetilde{\omega}_E - \omega)dt'\right] \quad (4d)$$

The quantities $COS(\theta/2)$ and $SIN(\theta/2)$ are complex functions that determine the partial contribution ("intensiveness") of real and virtual components to given NADS, whose asymptotic behavior at zero field and extremely strong fields is:

$$COS(\theta/2)\xrightarrow{E_0\to 0}1 \qquad , \qquad SIN(\theta/2)\xrightarrow{E_0\to 0}0 \quad (5a)$$

$$COS(\theta/2)\xrightarrow{E_0\to\infty}\sqrt{1/2} \qquad , \qquad SIN(\theta/2)\xrightarrow{E_0\to\infty}\sqrt{1/2} \quad (5b)$$

The virtual components of given NADS has "opposite" characteristics to these of real components of same NADS due to the electric dipole interaction (considered here) of the quantum system with the field and the associated selection rules in this case: $|\tilde{G}_r\rangle$ is $|g\rangle$-type of state while $|\tilde{G}_v\rangle$ is $|e\rangle$-type of state; $|\tilde{E}_r\rangle$ is $|e\rangle$-type of state while $|\tilde{E}_v\rangle$ is $|g\rangle$-type of state, Eq.(4). Consequently, *each NADS, ground $|\tilde{G}\rangle$ and excited $|\tilde{E}\rangle$, is a linear superposition of $|g\rangle$-type and $|e\rangle$-type of states of well-*defined physical coefficients, $COS(\theta/2)$ and $SIN(\theta/2)$, Eq.(3). In contrast to superposition (1), only one of the components in superposition (3) is real, $|g\rangle$ in $|\tilde{G}\rangle$ and $|e\rangle$ in $|\tilde{E}\rangle$, while the other component is virtual. The real components of NADS, $|\tilde{G}_r\rangle$ and $|\tilde{E}_r\rangle$, of energies (Bohr frequencies) $\tilde{\omega}_G$ and $\tilde{\omega}_E$, originate (after continuous evolution) from the respective BS, $|g\rangle$ and $|e\rangle$, of energies $\omega_g$ and $\omega_e$, respectively, Fig.2. The spectral width of the real components is not narrow due to broadening from the environment (damping) and the field, described by the imaginary parts of energies $\tilde{\omega}_G$ and $\tilde{\omega}_E$. The virtual components of NADS are *new states* that originate from the respective real components of NADS due to interaction of the quantum system with electromagnetic field. At zero field, the partial representation of the virtual components in the superposition (3) is zero, $SIN(\theta/2)=0$, Eq.(5a), and the NADS consist only of one real component – the initial BS from which it originates. Increasing the field amplitude, the partial representation of the virtual component increases while this of the real components decreases. At extremely strong fields, $E_0(t)\to\infty$, the intensiveness of the virtual and real components become equal - saturation of the virtual components, Eq.(5b). From a formal point of view, this behavior inspires the understanding that the creation of virtual component looks like "emission" of probability (to reside the quantum system on a given state) forced by the field from the real component to the respective virtual component of given NADS. From a point of view of the physical picture, the intensiveness of given component of NADS can be interpreted as the relative time that the quantum system resides on this component of the NADS. The virtual components of NADS, $|\tilde{G}_v\rangle$ and $|\tilde{E}_v\rangle$, of energies $\tilde{\omega}_G+\omega$ and $\tilde{\omega}_E-\omega$, originate from the respective real components of NADS, $|\tilde{G}_r\rangle$ and $|\tilde{E}_r\rangle$ of energies $\tilde{\omega}_G$ and $\tilde{\omega}_E$, respectively, after absorption/emission of one photon from/to the field of energy (frequency) $\omega$, Eq.(4), Fig.2. The spectral width of the virtual state is not narrow because the spectral width of the respective real states is not narrow and, also, the forcing electromagnetic field and thus, its photon spectrum, are not, in general, monochromatic. The field "lifts up"/"pulls down" the population, *e.g.*, the electron, by one photon energy (for single photon processes) from the real component of given NADS to the created virtual component of same NADS. Thus, in contrast to the real components, virtual components of given characteristics (depending on the type of transition – electric dipole, magnetic dipole, etc.) can be created *everywhere on the energy scale* depending on photon energy $\omega$. The virtual components, $|\tilde{G}_v\rangle$ and $|\tilde{E}_v\rangle$, cannot exist independently on the respective real components, $|\tilde{G}_r\rangle$ or $|\tilde{E}_r\rangle$, from which they originate, and the external electromagnetic field. The virtual components occur in the dressed state picture but not in the BS perturbation picture where the perturbations appear as corrections to the field-free BS and their energies.

We consider that the *virtual components* of NADS are *real physical states*, but not simply a mathematical construct, due to the following arguments. (*i*) Real population on the virtual components (like in any other real quantum state) has been observed experimentally [2, 3], see below; it may exceed by an order of magnitude the population of nearby real component; it can be transferred to populate the real components [2]. (*ii*) The real components "feel" the appearance of virtual components in their vicinity as the existence of any other real state. More particularly, real and virtual components of same symmetry repeal each other, Fig.2. Recall that $|\tilde{E}_r\rangle$ and $|\tilde{G}_v\rangle$ are $|e\rangle$-type of states while $|\tilde{G}_r\rangle$ and $|\tilde{E}_v\rangle$ are $|g\rangle$-type of states, Eq.4. The repulsion of states of same symmetry is well known for quantum systems with internal degrees of freedom (*e.g.*, diatomic molecules with internuclear distance being an internal degree of freedom) according to *the von Neumann-Wigner non-crossing rule* [5]. The repulsion of the components of NADS of same symmetry can be considered as a particular manifestation of the non-crossing rule for quantum systems without internal degrees of freedom (*e.g.*, atomic type systems considered here).

### 2.3. Multilevel nonadiabatic dressed states

Closed form solution of NADS exists in the two-level case. For the multilevel case, the ground and excited NADS are shown in Fig.3, extrapolating the two-level case by the following approach. For ground NADS, if we take the original ground BS $|g\rangle$ and repeat the derivation of NADS, as in [4], in a consecutive order with any one of the original excited BS $|e\rangle_1$, $|e\rangle_2$, $|e\rangle_3$ ..., the result will be generally same, as given by Eqs. (3) and (4). The difference is that at different detuning $\Delta\omega_i$ from the exact resonances, Fig.3, and dipole moment between states, the strength, $SIN(\theta/2)$ or $C_{vi}$, of the virtual state and the dynamic Stark shift will be different. If we now allow all excited BS to act simultaneously, as in the reality, each excited BS $|e\rangle_1$, $|e\rangle_2$, $|e\rangle_3$ ..., will evolve to the respective real component $|\tilde{E}_r\rangle_1$, $|\tilde{E}_r\rangle_2$, $|\tilde{E}_r\rangle_3$ ..., and, in addition, it will contribute to the creation of a respective virtual component $|\tilde{G}_v\rangle_1$, $|\tilde{G}_v\rangle_2$, $|\tilde{G}_v\rangle_3$ ... of the ground NADS, Fig.3. What is important, all virtual components, in contrast to the real components, have same energy equal to the energy of ground state real component $|\tilde{G}_r\rangle$ plus one photon energy, Fig.3. Eq.(6a) shows the evolution of the ground BS $|g\rangle$ toward the multilevel ground NADS $|\tilde{G}\rangle$, which consists of only one real state, $|\tilde{G}_r\rangle$, and a number of virtual states $|\tilde{G}_v\rangle_1$, $|\tilde{G}_v\rangle_2$, $|\tilde{G}_v\rangle_3$ ..., if a nonzero external field, $E_0 \neq 0$, is applied.

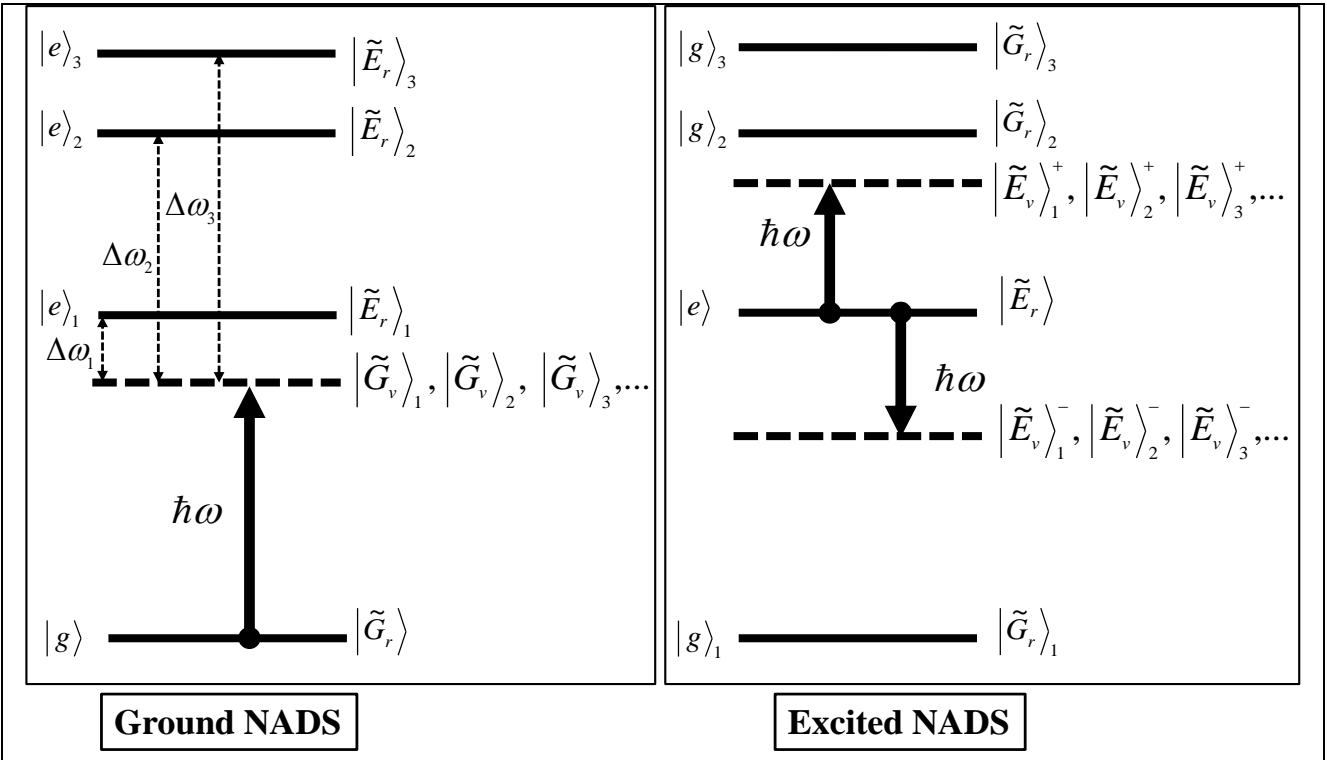

**Fig. 3.** *Energy structure of ground and excited multilevel NADS. The notations as* $|g\rangle_1$, $|g\rangle_2$, $|g\rangle_3$ ... *mean states of opposite parity with respect to that of the excited state* $|e\rangle$, *from which the process begins, rather than a notation for ground state.*

$$|g\rangle \xrightarrow{\quad E_0 \neq 0 \quad} |\tilde{G}\rangle = C_r |\tilde{G}_r\rangle + \sum_i C_{vi} |\tilde{G}_v\rangle_i \tag{6a}$$

$$|e\rangle \xrightarrow{\quad E_0 \neq 0 \quad} |\tilde{E}\rangle = C_r |\tilde{E}\rangle_r + \sum_i C_{vi}^+ |\tilde{E}_v\rangle_i^+ + \sum_i C_{vi}^- |\tilde{E}_v\rangle_i^- \tag{6b}$$

Similar structure has excited NADS, but upward $|\tilde{E}_v\rangle_i^+$ and downward $|\tilde{E}_v\rangle_i^-$ virtual components, Eq. (6b), can be created depending on the photon energy, Fig.3.

### 2.4. Experimental evidences and quantum superposition

There are some remarkable experiments [2, 3], from which, while having only nanosecond time resolution, one may distinguish the population dynamics of real and virtual states based on spectral resolution and different time behavior of the states. To get closer relation to such a work [2], we will apply our notations to the energy levels involved in [2], Fig.4. Atoms are excited adiabatically by a pump laser pulse at frequency $\omega$ from the real ground NADS $|\tilde{G}_r\rangle$ to the virtual ground NADS $|\tilde{G}_v\rangle$.

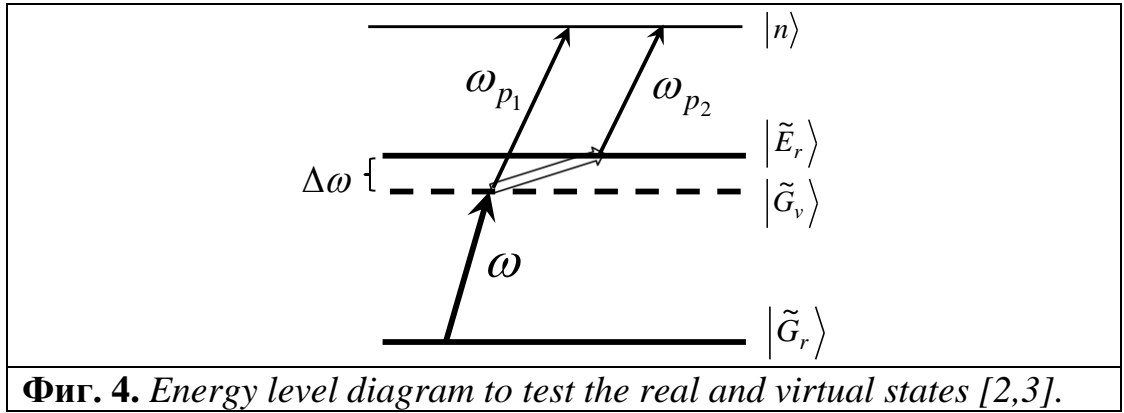

**Фиг. 4.** *Energy level diagram to test the real and virtual states [2,3].*

Two other laser fields at frequency $\omega_{p1}$ and $\omega_{p2}$ probe the population of virtual $|\tilde{G}_v\rangle$ and real $|\tilde{E}_r\rangle$ states, where $|n\rangle$ is a high-lying state. Although the field is adiabatic, population of real state $|\tilde{E}_r\rangle$ is observed *due to nonadiabatic factors from the environment* (collisions with other atoms, mainly). The population on the virtual state $|\tilde{G}_v\rangle$ is detected experimentally and it is an order of magnitude higher than that of the real state $|\tilde{E}_r\rangle$. The population of the virtual state is proportional to the intensity of the field and no population can be found on it once the field is switched off. The population on the real state $|\tilde{E}_r\rangle$ is proportional to the time integral from intensity of the field. One may conclude that, first, $|\tilde{G}_v\rangle$ state is physically populated following instantaneously the field and, next, part of this population is transferred to the real state $|\tilde{E}_r\rangle$. As any real state exists independently of the field, the population in $|\tilde{E}_r\rangle$ is trapped and accumulated there, leading to integral time dependence.

## 3. Reformulation of the quantum superposition principle

According to above considerations, the superposition in Eq.(1) is neither simultaneous nor coherent if the quantum states are eigenstates of adiabatically changing Hamiltonian, or, as it is more usual, the BS of a closed quantum system. In this relation, it is natural to find answers of the following questions: (*i*) "What kind of states take part in the quantum superposition?"; (*ii*) "What is the physical mechanism that can superimpose simultaneously two or more quantum states?"; (*iii*) "How to superimpose coherently quantum states if transitions between different eigenstates of adiabatic Hamiltonian result from the action of nonadiabatic factors, which are inherently stochastic, and this breaks the coherence?".

Above problems impose reformulation of the conventional superposition, Eq.(1). The superposition of states that meet above requirements are two-level NADS, Eqs.(3); multilevel NADS, Eqs.(6); or other states of similar features. These states, or superpositions, include only one initial real state (component) and one (two-level case) or a number (multilevel case) of virtual states (components). Each virtual state corresponds to a given real state that is coupled by allowed (in this case) electric dipole transition to the initial real state. If, in addition, magnetic dipole, electric quadrupole, etc. transitions are considered, more real states will be involve in the formation of virtual components, but with much smaller contribution. In any case, the superposition of all virtual states is instantaneous because they are created simultaneously with switching the electromagnetic field on. All virtual components, in the case of multilevel NADS, are coherent as they are created simultaneously with definite phases. In addition, as all virtual states have same energy (Bohr frequency), they oscillate with same frequency thus keeping this phase relation stable in time. According to this, the new formulation of the quantum superposition principle is:

Two-level case, Eqs. (3):

$$\left|\widetilde{G}\right\rangle = COS(\theta/2)\left|\widetilde{G}_r\right\rangle + SIN(\theta/2)\left|\widetilde{G}_v\right\rangle \tag{7a}$$

$$\left|\widetilde{E}\right\rangle = COS(\theta/2)\left|\widetilde{E}_r\right\rangle - SIN(\theta/2)\left|\widetilde{E}_v\right\rangle \tag{7b}$$

Multilevel case, Eqs. (6):

$$\left|\widetilde{G}\right\rangle = C_r\left|\widetilde{G}_r\right\rangle + \sum_i C_{vi}\left|\widetilde{G}_v\right\rangle_i \tag{8a}$$

$$\left|\widetilde{E}\right\rangle = C_r\left|\widetilde{E}\right\rangle_r + \sum_i C_{vi}^+\left|\widetilde{E}_v\right\rangle_i^+ + \sum_i C_{vi}^-\left|\widetilde{E}_v\right\rangle_i^- \tag{8b}$$

In all these formulations, the superposition of quantum states is a causal physical process of well-defined physical mechanism (*e.g.*, forced by electromagnetic field) but not an accidental linear combination of states.

## 4. Physical implications of the reformulated quantum superposition principle

Based on reformulated quantum superposition principle, some fundamental quantum mechanical problems as the collapse of wave function and the quantum measurement problem become understandable. Finally, inconsistency of the so called quantum "jumps" becomes clear from the way of formation of virtual states.

### 4.1 The quantum superposition and the collapse of the wave function

The collapse of the wave function takes place when the quantum system interacts with the environment or is subject to measurement. Then, the superposition wave function shrinks onto some of the eigenstates of the Hamiltonian [6], Eq. (9).

$$\sum_i c_i(t)\left|\psi(\vec{r}, t)\right\rangle_i \ \rightarrow \ \left|\psi(\vec{r}, t)\right\rangle_k \tag{9}$$

The collapse of the wave function cannot be described by the Schrödinger equation but only probabilistically, according to the Born's probability rule. The inability to explain "why the superposition of wave functions collapses onto only one of the superimposed quantum states" and "what is the physical mechanism behind it" is a fundamental problem, which dates since the early days of quantum mechanics.

Within the NADS picture, the creation of quantum superposition and collapse of wave function are simply two opposite directions of one and same physical process. Creation and collapse of quantum superposition for ground and excited NADS are:

Ground NADS, two-level case:

$$|g\rangle \xrightarrow{E_0 \neq 0} |\widetilde{G}\rangle = COS(\theta/2)|\widetilde{G}_r\rangle + SIN(\theta/2)|\widetilde{G}_v\rangle \tag{10a}$$

$$|\widetilde{G}\rangle = COS(\theta/2)|\widetilde{G}_r\rangle + SIN(\theta/2)|\widetilde{G}_v\rangle \xrightarrow{E_0 \to 0} |g\rangle \tag{10b}$$

Excited NADS, two-level case:

$$|e\rangle \xrightarrow{E_0 \neq 0} |\widetilde{E}\rangle = COS(\theta/2)|\widetilde{E}_r\rangle - SIN(\theta/2)|\widetilde{E}_v\rangle \tag{11a}$$

$$|\widetilde{E}\rangle = COS(\theta/2)|\widetilde{E}_r\rangle - SIN(\theta/2)|\widetilde{E}_v\rangle \xrightarrow{E_0 \to 0} |e\rangle \tag{11b}$$

Ground NADS, multilevel case:

$$|g\rangle \xrightarrow{E_0 \neq 0} |\widetilde{G}\rangle = C_r|\widetilde{G}_r\rangle + \sum_i C_{vi}|\widetilde{G}_v\rangle \tag{12a}$$

$$|\widetilde{G}\rangle = C_r|\widetilde{G}_r\rangle + \sum_i C_{vi}|\widetilde{G}_v\rangle_i \xrightarrow{E_0 \to 0} |g\rangle \tag{12b}$$

Excited NADS, multilevel case:

$$|e\rangle \xrightarrow{E_0 \neq 0} |\widetilde{E}\rangle = C_r|\widetilde{E}\rangle_r + \sum_i C_{vi}^+|\widetilde{E}_v\rangle_i^+ + \sum_i C_{vi}^-|\widetilde{E}_v\rangle_i^- \tag{13a}$$

$$|\widetilde{E}\rangle = C_r|\widetilde{E}_r\rangle + \sum_i C_{vi}^+|\widetilde{E}_v\rangle_i^+ + \sum_i C_{vi}^-|\widetilde{E}_v\rangle_i^- \xrightarrow{E_0 \to 0} |e\rangle \tag{13b}$$

The physical meaning of these equations is simple. All (…a)-equations represent the process of formation of quantum superposition, *i.e.*, NADS: when the electromagnetic field is switched on, the initial BS evolves toward a NADS, which consists of a single real state (into which the original BS evolves) and a number of virtual states. All (…b)-equations represent the process of collapse of the wave function: when the electromagnetic field is switching off, all virtual states disappear and the quantum superposition, *i.e.*, the NADS, collapses onto the single real state from which it originates. Thus, the collapse of the wave function is a direct consequence from the physical mechanism of creation of quantum superposition. This also shows that *the collapse of wave function is a real and causal physical process*, which is too fast (attoseconds or shorter) so as to be traced experimentally, for now.

### 4.2 The quantum superposition and the quantum measurement problem

The quantum measurement is a process of interaction of a quantum system with a measuring (usually, macroscopic) apparatus, or, in general, with the environment. The quantum system interacts with the measuring apparatus by some kind of field: electromagnetic field (our case), magnetic field (Stern-Gerlach apparatus), etc. Due to interaction with electromagnetic field, the states of the quantum system evolve from the initial BS to the respective NADS and form entangled state with the respective states of the field of the apparatus. The quantum measurement is closely related to the collapse of the wave function onto one only state of the quantum system [6], which has been already explained above. Another main point in the quantum measurement is to explain the seemingly random probabilistic outcome of the measurement.

The central part in the evolution of quantum states, Fig.2, is nonadiabatic loop, shown in Fig.5. The nonadiabatic loop consists of two arms, ground NADS arm (left, as chosen here) and excited NADS arm (right). The ground and excited NADS are adiabatically decoupled at lack of nonadiabatic factors. In this case, if the quantum system is initially in the ground BS $|g\rangle$, it will evolve toward ground NADS $|\tilde{G}\rangle$ and it will remain in that state with no transition to the excited NADS $|\tilde{E}\rangle$. Due to action of nonadiabatic factors, quantum transition from $|\tilde{G}\rangle$ to $|\tilde{E}\rangle$, more particularly, from $|\tilde{G}_v\rangle$ to $|\tilde{E}_r\rangle$ (due to minimal energy gap $\Delta E$), will occur. In fact, the quantum system is trapped in $|\tilde{E}_r\rangle$, as it is stable state (within the lifetime), and the population can be accumulated in this state, as also the experiment shows [2]. Trapping the quantum system in $|\tilde{E}_r\rangle$ terminates its evolution within the ground NADS. If the action of the electromagnetic field continue, it will start forming a virtual state $|\tilde{E}_v\rangle$ from $|\tilde{E}_r\rangle$ and the quantum system will evolve within the excited NADS $|\tilde{E}\rangle$. The quantum system will remain in $|\tilde{E}\rangle$ until the next action of nonadiabatic factors transfer it from $|\tilde{E}\rangle$ to $|\tilde{G}\rangle$, more particularly, from $|\tilde{E}_v\rangle$ to $|\tilde{G}_r\rangle$ (due to minimal energy gap $\Delta E$). This terminates the evolution of the quantum system within the excited NADS and closes the nonadiabatic loop. Depending on the duration of the physical processes involved in the loop and in the measurement, the measurement may lasted from only part of the loop up to a number of loops. As the nonadiabatic factors have, in general, stochastic character, transition from one to other nonadiabatic arm and, thus, from one NADS to other NADS, have same stochastic character. Thus, *the nonadiabatic physical processes within the nonadiabatic loop are responsible for the random outcome of the measurement process*. The particular outcome of the measurement depends on which state the quantum system is in at the "end of the measurement". As "end of measurement" (depending on the measurement scheme) we will understand the instant of time at which: the field is switched off, the quantum system leaves the area where the field is localized, some auxiliary field move the quantum system to a state out of the loop, etc. Thus, if at the end of measurement the quantum system is in the ground NADS $|\tilde{G}\rangle$, it will be found in the ground BS $|g\rangle$ and energy $E_g$ will be measured. If at the end of measurement the quantum system is in the excited NADS $|\tilde{E}\rangle$, it will be found in the excited BS $|e\rangle$ and energy $E_e$ will be measured.

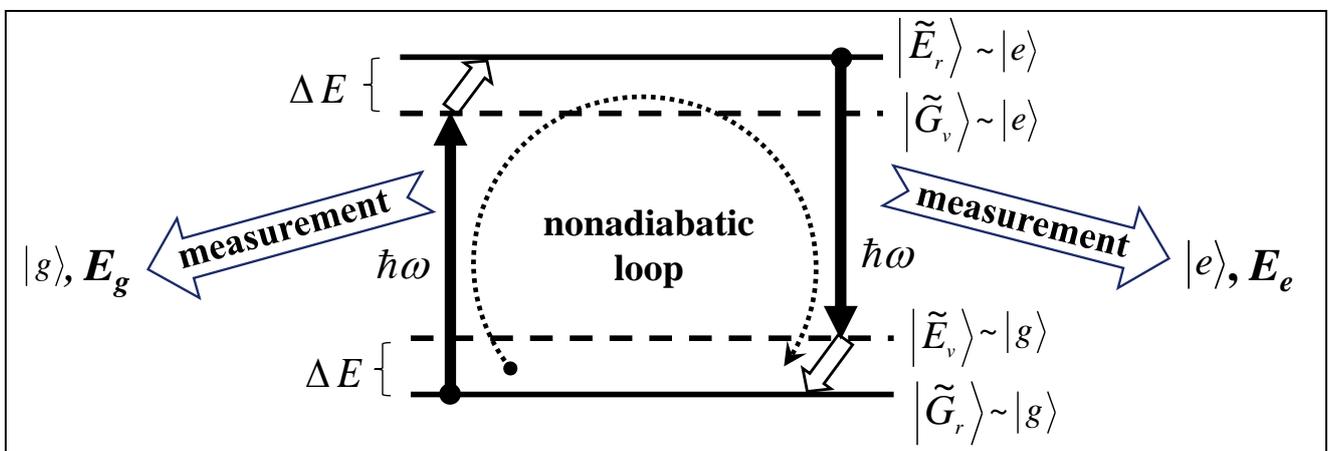

**Fig. 5.** *Nonadiabatic loop and the random outcome of the quantum measurement*

### 4.3 Quantum "jumps"

The above considerations automatically explain the problem with the so called *quantum "jumps"*. The later takes place for *isolated quantum systems*, for which all intermediate states, apart from its eigenstates, are forbidden. For *open quantum systems*, *e.g.*, at presence of external electromagnetic field, the system can be excited (with different probability) to any intermediate state, *virtual state*, depending on photon energy. Thus, there is no place for quantum jumps in the dressed state picture.

### 5. Conclusion

The quantum superposition principle is reformulated. The collapse of the wave function and the quantum measurement problem are explained on that ground.